\documentclass[aps,preprint]{revtex4}%
\usepackage{amsfonts}
\usepackage{amsmath}
\usepackage{amssymb}
\usepackage{graphicx}%
\providecommand{\U}[1]{\protect\rule{.1in}{.1in}}
\newtheorem{theorem}{Theorem}
\newtheorem{acknowledgement}[theorem]{Acknowledgement}

\begin{document}
\title{Rydberg states of hydrogen-like ions in braneworld}
\author{F. Dahia}
\author{E. Maciel}
\author{A. S. Lemos}
\affiliation{Department of Physics, Federal University of Para\'{\i}ba - Jo\~{a}o Pessoa
-PB - Brazil }
\pacs{PACS number}

\begin{abstract}
It has been argued that precise measurements of optical transition frequencies
between Rydberg states of hydrogen-like ions could be used to obtain an
improved value of the Rydberg constant and even to test Quantum
Electrodynamics (QED) theory more accurately, by avoiding the uncertainties
about the proton radius. Motivated by this perspective, we investigate the
influence of the gravitational interaction on the energy levels of
Hydrogen-like ions in Rydberg states within the context of the braneworld
models. As it is known, in this scenario, the gravitational interaction is
amplified in short distances. We show that, for Rydberg states, the main
contribution for the gravitational potential energy does not come from the
rest energy concentrated on the nucleus but from the energy of the
electromagnetic field created by its electrical charge, which is spread in
space. The reason is connected to the fact that, when the ion is in a Rydberg
state with high angular momentum, the gravitational potential energy is not
computable in zero-width brane approximation due to the gravitational
influence of the electrovacuum in which the lepton is moving. Considering a
thick brane scenario, we calculate the gravitational potential energy
associated to the nucleus charge in terms of the confinement parameter of the
electric field in the brane. We show that the gravitational effects on the
energy levels of a Rydberg state can be amplified by the extra dimensions even
when the compactification scale of the hidden dimensions is shorter than the
Bohr radius.

\end{abstract}
\maketitle

\section{Introduction}

In the past decades we have seen a renewed interest in extra-dimensional
theories stimulated by the braneworld scenario, according to which our
ordinary universe is a submanifold embedded in a higher-dimensional space
\cite{ADD1,ADD2,RS1,RS2}. The fundamental characteristic of these models is
the assumption that matter and fields are confined in a three-dimensional
space (the 3-brane), while gravity has access to the whole space, implying
that the gravitational field could feel direct effects of the extra dimensions
on a length scale much greater than the scale where the standard model fields
are directly affected. Since the gravitational interaction has been tested in
short distance only recently, then braneworld scenario may allow the
formulation of phenomenologically viable models in which the hidden dimensions
can be much larger than the four-dimensional Planck length $\left(  10^{-35}%
\operatorname{m}%
\right)  $, which is the characteristic size of the fifth direction postulated
by the Kaluza-Klein model, the first modern higher-dimensional theory.

Direct laboratory tests of the inverse square law impose that the
compactification radius $(R)$ should be less than 44$%
\operatorname{\mu m}%
$ \cite{Hoyle2007,review}, for instance. This is the most stringent constraint
when the supplementary space has only one extra dimension. Regarding greater
codimensions, experimental limits from astrophysics and colliders are tighter,
but, according to current data, they are still greater than Planck length by
15 orders at least
\cite{SN,neutronstar,colliders,lhc,monojet,landsberg,pdgAlex}.

The original purpose of braneworld theories was to explain the hierarchy
problem, i.e., why gravity is so much weaker in comparison to the other
fundamental forces \cite{ADD1,ADD2,RS1,RS2}. According to these models, the
reason, roughly speaking, is that gravity is the only field that spreads in
every direction, hence it seems weaker than the other interactions in the long
distance scale, where the usual four-dimensional behavior is recovered.

On the other hand, in distances smaller than the size of the extra dimensions,
theses models predict that the gravitational force is amplified in comparison
to the Newtonian three-dimensional force by a factor that depends on the
number and on the size of the additional dimensions. This fact has motivated
many authors to study the effects of the gravitational field in the atomic and
molecular system as a way to obtain independent experimental bounds for
parameters of the higher-dimensional theories
\cite{specforxdim1,atomicspec,Li,Wang,specforxdim2,molecule,dahia}.

Here we intend to study the influence of the extra dimensions on Rydberg
states of hydrogen-like ions, motivated by recent developments in the
spectroscopy area which suggest that precision measurements of optical
transitions between Rydberg states can become an efficient method to test QED
theory and also to determine the value of the Rydberg constant more accurately
\cite{Jentschura}.

The current uncertainty of Rydberg constant is of the order of $10^{-12}$
\cite{codata} and its value is extracted from the comparison of theoretical
predictions and measurements of transition frequencies in hydrogen and
deuteron, which includes transitions between $S$-states
\cite{codata,Jentschura}. It happens that, regarding $S$-states, theoretical
calculations are mainly limited by the uncertainty about the proton charge
radius\cite{Jentschura}, whose value is also a source of a controversy known
as radius proton puzzle \cite{nature,science,carl,krauth}, which is a
discrepancy between measurements of the proton size as inferred from the
muonic hydrogen spectroscopy and the CODATA value \cite{codata}, based on the
proton-electron interaction.

In Rydberg states with high angular momentum, as the influence of the internal
structure of the nucleus is negligible, the predictions are practically
independent of the proton radius and then measurements would be free from
those uncertainties. With recent calculations of QED corrections for
higher-quantum number states, the accuracy of the predicted energy level could
reach parts in 10$^{17}$ \cite{Jentschura}. At the same time, advances on
experimental techniques, such as optical frequency combs \cite{opticalcomb},
promises to reduce the relative uncertainties to the order of $10^{-19}$ in
measurements of transition frequencies around the optical range
\cite{opticalmetrology}. Together, these theoretical and experimental advances
may allow precise measurements of certain optical transitions between
appropriate Rydberg states which, in principle, could lead to an improved
value of the Rydberg constant \cite{Jentschura}.

Faced with such expectations, it seems relevant to investigate as how the
extra dimensions could interfere in the energy level of the Rydberg states in
the braneworld scenario. The effects of gravitational interaction on the
$S$-states have already been considered before \cite{dahia}. An interesting
point to mention here is that when the space has more than two extra
dimensions, the mean gravitational potential energy of a hydrogen-like atom in
a $S$-state, $\left\langle U\right\rangle _{S}$, diverges if the brane
structure is not taken into account. Indeed, this is connected to fact that
the gravitational potential, $\varphi$, produced by the proton mass is not
computable in the interior of the nucleus in the approximation of zero-width
brane \cite{colliders,dahia,effbrane}.

On the other hand, when the atom is in a Rydberg state the situation may be
different. For a state with a high enough angular momentum, the leading term
of $\left\langle U\right\rangle _{Ry}$ is finite in the thin brane limit, so
the zero-width brane is a valid idealization. However, under such
circumstances, the amplification of the gravitational potential energy by
extra dimensions is significant only if the compactification radius is greater
than the atomic Bohr radius. Thus, considering the current constraints on $R$
\cite{pdgAlex}, at the first sight, we could be led to think that extra
dimensions would have little influence on Rydberg states.

Nevertheless, it is important to realize that the geometry around the nucleus
should be similar to the Reissner-Nordstrom geometry, due to the nucleus
electric charge. In a Reissner-Nordstrom spacetime, besides the gravitational
potential, $\varphi$, produced by the rest mass, there is also the
gravitational influence associated to the energy of the electromagnetic field
created by the electric charge of the source, which, in the weak field regime,
can be described by a certain potential $\chi$.

Although, in the traditional three-dimensional picture, $\chi$ is smaller than
the potential $\varphi$, this relation in the braneworld scenario can be
different in a certain region outside the nucleus. As the electromagnetic
field is spread in the three-dimensional space, the potential $\chi$ diverges
in every point of the brane, in the zero-thickness limit. Thus, in order to
compute $\chi$, the distribution of the electric energy inside the brane
should be considered. Addressing this problem in the thick brane scenario, we
study the Green function associated to the gravitational potential in length
scales smaller than $R$, where the effects of extra dimensions are stronger.
With this approximated Green function, we determine the short-distance
potential $\chi_{s}$. We find that it can be greater than the short-distance
potential $\varphi_{s}$ outside the nucleus. This can lead to interesting
consequences. Indeed, considering a Hydrogen-like ion in a Rydberg state, we
show that the extra dimensions can amplify the gravitational potential energy
of the ion even when the compactification scale $R$ is smaller than the Bohr
radius, due to the behavior of $\chi_{s}$.

Finally, we estimate the effects of higher-dimensional gravity on particular
optical transitions of  Hydrogen-like ions and discuss the possibility of
using the spectroscopy of Rydberg states in the search for hidden dimensions.

\section{The gravitational field produced by the nucleus}

In the original ADD-braneworld model \cite{ADD1}, the spacetime has $\delta$
additional spacelike dimensions with the topology of a torus $T^{\delta}$. In
the ground state of the model, the extra dimensions have a certain radius $R$,
and it is assumed that the metric is flat. This means that the energy of the
brane itself does not curve the bulk at long distances compared to the length
scale of the brane, by the intervention of some mechanism \cite{ADD1,ADD2}. It
is also admitted that the gravitational field produced by the matter confined
in the brane is governed by the Einstein-Hilbert action extended to higher
dimensions:%
\begin{equation}
S_{G}=\frac{c^{3}}{16\pi G_{D}}\int d^{4}xd^{\delta}z\sqrt{-\hat{g}%
}\mathcal{\hat{R}}, \label{S}%
\end{equation}
where $\mathcal{\hat{R}}$ is the scalar curvature of the bulk, $\hat{g}$ is
the determinant of the metric whose signature is assumed to be $(-,+,...,+)$
and $G_{D}$ is the gravitational constant defined in the higher-dimensional
space. The coordinates $x$ and $z$ refer to parallel and transversal
directions with respect to the brane. Due to the topology of the supplementary
space, the metric is periodic in the extra directions and can be expanded in a
Fourier series with respect to the $z$-coordinates, giving rise to the
so-called KK-modes. The zero-mode is supposed to reproduce the
four-dimensional gravitational field at large distance in comparison to $R$.
This condition requires that $G_{D}$ should be related to the Newtonian
gravitational constant $G$ according to the following formula
\cite{ADD1,colliders}:%
\begin{equation}
G_{D}=G\left(  2\pi R\right)  ^{\delta}. \label{GD}%
\end{equation}
Additionally, in order to get the correct Newtonian limit, it is also
necessary that some mechanism ensures the stabilization of the volume of the
supplementary space at large distance \cite{antoniadis}.

The extremization of the action (\ref{S}) yields the higher-dimensional
version of the Einstein equations. In the weak field regime, which we assume
to be valid in the atomic domain, the metric can be written as $g_{AB}%
=\eta_{AB}+h_{AB}$. Here the capital Latin indices run from $0$ to $3+\delta$,
$\eta_{AB}$ is the Minkowski metric and $h_{AB}$ is a small perturbation of
the order of $G_{D}M$. In a coordinate system in which the gauge condition
\begin{equation}
\partial_{A}\left(  h^{AB}-\frac{1}{2}\eta^{AB}h_{C}^{C}\right)  =0
\label{gauge}%
\end{equation}
is satisfied, the linearized equations reduce to the form
\begin{equation}
\square h_{AB}=-\frac{16\pi G_{D}}{c^{4}}\bar{T}_{AB}, \label{linear eq}%
\end{equation}
where $\square$ is the D'Alembertian operator associated to the Minkowski
metric and $\bar{T}_{AB}=\left[  T_{AB}-(\delta+2)^{-1}\eta_{AB}T_{C}%
^{C}\right]  $ is defined in terms of the energy-momentum tensor $T_{AB}$ of
the source.

Considering the topology $%
\mathbb{R}
^{3}\times T^{\delta}$, the solution of equation (\ref{linear eq}) for static
sources is
\begin{equation}
h_{AB}\left(  \vec{X}\right)  =\frac{16\pi G_{D}\Gamma(\frac{\delta+3}{2}%
)}{(\delta+1)2\pi^{\left(  \delta+3\right)  /2}c^{4}}\sum\limits_{i}\left(
\int\frac{\bar{T}_{AB}\left(  \vec{X}^{\prime}\right)  }{\left\vert \vec
{X}-(\vec{X}^{\prime}+\vec{K}_{i})\right\vert ^{1+\delta}}d^{3+\delta
}X^{\prime}\right)  , \label{solution}%
\end{equation}
where $\vec{X}=\left(  \vec{x},\vec{z}\right)  $, $\vec{K}_{i}=$ $2\pi
R\left(  0,0,0,k_{1},..,k_{\delta}\right)  $ and each $k_{i}$ is an integer
number. If we consider $T^{\delta}$ as a manifold embedded in $%
\mathbb{R}
^{\delta}$, then the vectors $\vec{K}_{i}$ may be viewed as the localization
of the mirror images of the source induced by the topology of the
supplementary space on $%
\mathbb{R}
^{\delta}$ (the covering space). The presence of the mirror images makes the
solution periodic with respect to the $z$-coordinates as demanded by the
topology. It can be shown that the higher-dimensional Green function recovers
the four-dimensional behavior for long distances $\left\vert \vec
{x}\right\vert >>R$ \cite{kehagias}. On the other hand, in short distance (
$\left\vert \vec{x}-\vec{x}^{\prime}\right\vert <R$ ), the Green function is
dominated by the first term of the expansion (\ref{solution}). In this paper,
we are interested in studying the effects of the short-distance behavior of
the gravitational interaction on the atomic energy levels. Thus for the sake
of simplicity, we are going to take only the first term $(\vec{K}_{0}=0)$ of
the series (\ref{solution}) as an approximation of the solution. Of course,
this approximation gives a low estimate of the gravitational effect, since all
terms of the series (\ref{solution}) we are neglecting have the same sign.

In the context of the braneworld, it is assumed that, for a length scale much
greater than the brane thickness, the energy-momentum tensor of the confined
fields has the form \cite{colliders}:%
\begin{equation}
T_{AB}\left(  x,z\right)  =\eta_{A}^{\mu}\eta_{B}^{\nu}T_{\mu\nu}\left(
x\right)  f\left(  z\right)  ,
\end{equation}
where $T_{\mu\nu}\left(  x\right)  $ is the ordinary energy-momentum tensor of
the four-dimensional fields that live in the brane and $f\left(  z\right)  $
is a normalized distribution very concentrated around the brane, which is
approximately given by a delta-like distribution in the thin brane limit. The
function $f(z)$ describes the confinement of a field in the brane and, in
principle, it could be a different function for each type of field.

In our system, the atomic nucleus is the source of the gravitational field.
Due to its electric charge, it is reasonable to expect that the spacetime
geometry around the nucleus should be similar to the Reissner-Nordstrom
geometry. Based on this, it is convenient to decompose the energy-momentum
tensor as a sum of two terms: $T_{\mu\nu}=T_{\mu\nu}^{\left(  0\right)
}+T_{\mu\nu}^{\left(  EM\right)  }$, where $T_{\mu\nu}^{\left(  0\right)  }$
describes the rest energy concentrated inside the nucleus and $T_{\mu\nu
}^{\left(  EM\right)  }$ represents the stress-energy tensor of the
electromagnetic field created by the charge, which is spread in the space.

We deal with each term separately. The first tensor can be written as
$T_{AB}=c^{2}\rho\eta_{A}^{0}\eta_{B}^{0}$, where $\rho$ is the mass density
of the source. On its turn, in SI units, $T_{\mu\nu}^{\left(  EM\right)
}=\epsilon_{0}c^{2}\left(  F_{\mu\lambda}F_{\nu}^{\;\lambda}-\frac{1}{4}%
\eta_{\mu\nu}F_{\alpha\beta}F^{\alpha\beta}\right)  $, where $F_{\mu\nu}$ is
the electromagnetic tensor and $\epsilon_{0}$ is the electric permittivity of
the free space. In a first approach, let us ignore the contribution of the
magnetic field produced by the proton's magnetic dipole. In this
approximation, only the components of the electric field, $F_{0i}=E_{i}/c$,
are non-null (in the zero order of $G_{D}$). Take into account all this
consideration, we can show that the metric can be written as:%

\begin{align}
ds^{2}  &  =-\left(  1+\frac{2}{c^{2}}\varphi_{s}+\frac{2(2+\delta)}%
{c^{2}(1+\delta)}\chi_{s}\right)  \left(  dx^{0}\right)  ^{2}+\nonumber\\
&  \left(  1-\frac{2}{c^{2}(1+\delta)}\varphi_{s}\right)  \left[
(1+\frac{2(2+\delta)}{c^{2}(1+\delta)}\lambda_{1,s})dr^{2}+(1+\frac
{2(2+\delta)}{c^{2}(1+\delta)}\lambda_{2,s})r^{2}(d\theta^{2}+\sin^{2}\theta
d\phi^{2})\right] \nonumber\\
&  +\left(  1-\frac{2}{c^{2}(1+\delta)}\varphi_{s}\right)  d\vec{z}^{2},
\label{Dmetric}%
\end{align}
where $x^{0}=ct$, the coordinates $\left(  r,\theta,\phi\right)  $ are the
usual spherical coordinates associated to the "almost Cartesian" coordinates
$(x^{1},x^{2},x^{3})$, defined by eq. (\ref{gauge}). The function $\varphi
_{s}$ plays the role of a Newtonian potential produced by the nuclear mass in
$%
\mathbb{R}
^{3+\delta}$:
\begin{equation}
\varphi_{s}\left(  \vec{X}\right)  =-\hat{G}_{D}\int\frac{\rho\left(  \vec
{x}^{\prime}\right)  f_{m}(z)}{\left\vert \vec{X}-\vec{X}^{\prime}\right\vert
^{1+\delta}}d^{3+\delta}X^{\prime}, \label{phi}%
\end{equation}
where $\hat{G}_{D}=4G_{D}\Gamma(\frac{3+\delta}{2})/[(2+\delta)\pi^{\left(
1+\delta\right)  /2}].$ The sub-index $s$ emphasizes that $\varphi_{s}$ is the
potential given by the short-distance Green function, which corresponds to
first term of the series (\ref{solution}). Analogously, $\chi_{s}$ is the
short-distance gravitational potential produced by the energy of the
electromagnetic field $u=(\epsilon_{0}E^{2}/2)$ created by the electric
charge:%
\begin{equation}
\chi_{s}\left(  \vec{X}\right)  =-\frac{\hat{G}_{D}}{c^{2}}\int\frac{u\left(
\vec{x}^{\prime}\right)  f_{e}\left(  z\right)  }{\left\vert \vec{X}-\vec
{X}^{\prime}\right\vert ^{1+\delta}}d^{3+\delta}X^{\prime}. \label{chi}%
\end{equation}
The spatial components, $T_{ij}^{(EM)}$, of the electromagnetic stress-energy
tensor, give rise to the functions $\lambda_{2,s}$ and $\lambda_{1,s}$, which
are defined by:%
\begin{align}
\lambda_{2,s}  &  =-\chi_{s}-\pi\frac{\hat{G}_{D}}{c^{2}}\int\frac
{\epsilon_{0}E^{2}r^{\prime2}\left(  \sin^{3}\theta\right)  dr^{\prime}%
d\theta}{\left\vert \left(  r^{2}+r^{\prime2}-2rr^{^{\prime}}\cos
\theta\right)  +\left\vert \vec{z}\right\vert ^{2}\right\vert ^{\frac
{1+\delta}{2}}}f_{e}(z)d^{\delta}z^{\prime},\label{alfa}\\
\lambda_{1,s}  &  =-\chi_{s}-2\pi\frac{\hat{G}_{D}}{c^{2}}\int\frac
{\epsilon_{0}E^{2}r^{\prime2}\left(  \cos^{2}\theta\sin\theta\right)
dr^{\prime}d\theta}{\left\vert \left(  r^{2}+r^{\prime2}-2rr^{^{\prime}}%
\cos\theta\right)  +\left\vert \vec{z}\right\vert ^{2}\right\vert
^{\frac{1+\delta}{2}}}f_{e}(z)d^{\delta}z^{\prime}. \label{beta}%
\end{align}

As we shall see in the next section, in the gravitational sector of the atomic
Hamiltonian $\left(  H_{G}\right)  $, the leading term comes from the
component $g_{00}$, which depends on the potential $\varphi_{s}$ and $\chi
_{s}$. To determine explicitly the functions $\varphi_{s}$ and $\chi_{s}$, it
is important to look at the internal structure of the brane to see how the
fields are localized inside. In a field-theoretic framework, topological
defects are possible realizations of a brane, since these structures are
capable of localizing fermions in their cores, as illustrated in Ref.
\cite{rubakov}, where it is shown that Dirac fields can be trapped inside a
domain wall by means of a Yukawa-like interaction. In this context, usually
known as thick brane scenario, a delta-like localization is replaced by a
non-singular confinement where the fermion's states are described by a regular
wave-function with a tiny width $\sigma$ in the transversal directions.

Following these ideas, in Ref. \cite{dahia}, we calculated the gravitational
potential, $\varphi_{s}$, produced by the proton mass $M_{p}$ in the thick
brane scenario. By admitting that the proton wave-function in the transversal
directions has a Gaussian profile, we have estimated the influence of the
gravitational interaction on the energy of $S$-states of a Hydrogen atom. The
leading contribution is proportional to $\hat{G}_{D}mM_{p}/a_{0}^{3}%
\sigma^{\delta-2}$, where $m$ is the electron mass and $a_{0}$ is the Bohr
radius. It is clear, from this expression, that, the calculation diverges,
even at the tree level, in the thin brane limit. Therefore, the mass
distribution of the nucleus inside the brane cannot be neglected when the atom
is in $S$-states. We should highlight that the mentioned term corresponds to
the mean value of the gravitational potential energy integrated in the
interior of the nucleus, which we shall denote as $\left\langle H_{G}%
\right\rangle _{in}$. On its turn, outside the nucleus, the dominant
contribution is proportional to $\hat{G}_{D}Mm/a_{0}^{1+\delta}$, if the
compactification scale is greater than the Bohr radius. As, in realistic
scenarios, $\sigma<<a_{0}$ then the interior contribution $\left\langle
H_{G}\right\rangle _{in}$ is much greater than the exterior contribution
$\left\langle H_{G}\right\rangle _{out}$ for the energy of $S$-states.

However, if the atom is in a Rydberg states with large angular momentum, the
inverse happens. Indeed, in states with angular momentum $l$, the internal
contribution is reduced by a factor of the order of $(r_{N}/a_{0})^{2l}$,
where $r_{N}$ is the radius of the nucleus. Therefore, for some angular
momentum higher than%
\begin{equation}
2l>\left(  \delta-2\right)  \frac{\ln\left(  a_{0}/\sigma\right)  }{\ln\left(
a_{0}/r_{N}\right)  },
\end{equation}
the exterior contribution $\left\langle H_{G}\right\rangle _{out}$ become
greater than $\left\langle H_{G}\right\rangle _{in}$. Therefore, as the
leading term does not depend on the brane thickness, we may say that a
zero-width brane is a valid approximation for such states with high quantum
numbers. Replacing $f_{m}(z)$ in equation (\ref{phi}) by a delta-Dirac
distribution, we find that, in the exterior region, the short-distance
gravitational potential produced by the nucleus mass is
\begin{equation}
\varphi_{s}=-\hat{G}_{D}\frac{M}{r^{1+\delta}}. \label{shortphi}%
\end{equation}
With respect to the complete potential given by the series (\ref{solution}),
the term (\ref{shortphi}) is the leading contribution in the region
$r_{N}<<r<<R$. By using (\ref{GD}), we can conclude that the contribution of
(\ref{shortphi}) to the gravitational potential energy of the atom in a
Rydberg state will be of the order of $\left(  GMm/a_{0}\right)
(R/a_{0})^{\delta}$ and, therefore, extra dimensions would amplify
significantly the gravitational energy of the atom in a Rydberg state only if
$R>>a_{0}$.

This conclusion is not necessarily valid when we consider the potential
$\chi_{s}$ produced by the electromagnetic energy. If $f_{e}(z)$ is
approximated by a delta-like distribution in equation (\ref{chi}), the
potential $\chi_{s}$ diverges everywhere, not only inside the nucleus, as
happens with the potential $\varphi_{s}$, since the electromagnetic is spread
in the space. Therefore, due to the behavior of potential $\chi_{s}$, the
zero-width brane is not a valid idealization even when we are dealing with
ions in Rydberg states.

Thus, in order to compute $\chi_{s}$, we have to consider the distribution of
the electric energy inside the thick brane. With the purpose of obtaining some
estimates, we are going to admit that the electric energy is uniformly
distributed inside a compact region of the brane with a size $\varepsilon$,
which may have the same order of the brane thickness. Thus, if $V_{\delta
}\left(  \varepsilon\right)  $ denotes the volume of a ball with a radius
$\varepsilon$ in the supplementary space, $f_{e}(z)$ can be defined as the
step function:%
\begin{equation}
f_{e}(z)=%
\genfrac{\{}{.}{0pt}{}{1/V_{\delta}\left(  \varepsilon\right)  ,\hspace
{0.15in}z^{2}\leq\varepsilon,}{0,\hspace{0.15in}z^{2}>\varepsilon.}
\label{f}%
\end{equation}
Taking this function in the expression (\ref{chi}) and integrating it with
respect to the angular coordinates, the potential $\chi_{s}$ evaluated in the
brane ( $\vec{z}=0$ ) can be written as $\chi_{s}=\chi_{+}-\chi_{-}$ , where%
\begin{equation}
\chi_{\pm}\left(  r\right)  =\frac{\hat{G}_{D}S_{\delta}}{V_{\delta}\left(
\varepsilon\right)  }\frac{2\pi}{(\delta-1)}\frac{1}{r}\int\frac{u\left(
r^{\prime}\right)  }{\left[  (r\pm r^{\prime})^{2}+z^{\prime2}\right]
^{\frac{-1+\delta}{2}}}\left(  z^{\prime}\right)  ^{\delta-1}r^{\prime
}dr^{\prime}dz^{^{\prime}}. \label{chi+-}%
\end{equation}
Here $S_{\delta}$ is the hyper-area of a spherical hypersurface of $\delta$
dimensions with a unit radius. In (\ref{chi+-}), the integration interval of
the transversal variable is $0<z^{\prime}<\varepsilon$.

The divergent term in the thin brane limit comes from the function $\chi_{-}$
as the integrating variable $r^{\prime}$ passes through $r$. In the context of
a thick brane scenario, this "dangerous" term can be isolated and calculated
explicitly. As we are interested in studying Rydberg states with high angular
momentum, we may restrict our analysis to points far from the nucleus ( $r>>$
$r_{N}$). In the calculation of $\chi_{s}$, it is convenient to separate the
integration domain in two parts: the region inside the interval $\left\vert
r^{\prime}-r\right\vert \leq R,$ where the short-distance behavior of the
Green function is dominant, and the external region $\Omega$. Noticing that
the zero-thick brane approximation is valid in $\Omega$, we can write:
\begin{align}
\chi_{s}\left(  r\right)   &  \approx\frac{2\pi\hat{G}_{D}}{(\delta-1)}%
\frac{1}{r}\int_{\Omega}\left[  \frac{1}{\left\vert r+r^{\prime}\right\vert
^{-1+\delta}}-\frac{1}{\left\vert r-r^{\prime}\right\vert ^{-1+\delta}%
}\right]  u\left(  r^{\prime}\right)  r^{\prime}dr^{\prime}\nonumber\\
&  -\frac{\hat{G}_{D}S_{\delta}}{V_{\delta}\left(  \varepsilon\right)  }%
\frac{2\pi}{(\delta-1)}\frac{1}{r}\underset{\left\vert r^{\prime}-r\right\vert
\leq R}{\int}\frac{u\left(  r^{\prime}\right)  }{\left[  (r\pm r^{\prime}%
)^{2}+z^{\prime2}\right]  ^{\frac{-1+\delta}{2}}}\left(  z^{\prime}\right)
^{\delta-1}r^{\prime}dr^{\prime}dz^{^{\prime}}. \label{chi u}%
\end{align}

To proceed further we need now to specify the electromagnetic energy
distribution in the three-dimensional space by means of the function $u$. To
be consistent, the model should reproduce the ordinary behavior of
electromagnetic field in a length scale greater than the brane thickness
($r>>\varepsilon)$. It is important to remark at this point that, the
compactification scale $\left(  R\right)  $ is admitted to be greater than
$\varepsilon$ in our approach, so the gravitational field is the only field
directly affected by extra dimensions in the region $r>>\varepsilon$. Based on
this considerations, we will assume that in the three-dimensional space the
energy density of the electric field has the usual form given by:
\begin{equation}
u=\frac{1}{2}\epsilon_{0}E^{2}=\frac{Q^{2}}{32\pi^{2}\epsilon_{0}}\times%
\genfrac{\{}{.}{0pt}{}{\frac{r^{\prime2}}{r_{N}^{6}},r^{\prime}\leq
r_{N}}{\frac{1}{r^{\prime4}},r^{\prime}\geq r_{N}}%
\label{u}%
\end{equation}
which is the well-known energy of electromagnetic field produced by a charge
$Q$ uniformly distributed inside a ball of radius $r_{N}$ in three-dimensional
space. With this energy distribution, we find:%

\begin{equation}
\chi_{s}=-\frac{\beta_{\delta}}{16\pi\epsilon_{0}c^{2}}\frac{\hat{G}_{D}Q^{2}%
}{\varepsilon^{\delta-2}r^{4}}+\mathcal{O}_{1}+\mathcal{O}_{2},
\label{chi sol}%
\end{equation}
where $\beta_{\delta}$ can be written in terms of the gamma function as:%
\begin{equation}
\beta_{\delta}=\frac{\delta}{(\delta-1)}\frac{\sqrt{\pi}\Gamma\left(
\frac{\delta-2}{2}\right)  }{2\Gamma\left(  \frac{\delta-1}{2}\right)  },
\end{equation}
and the corrections terms are of the following order:%
\begin{align*}
\mathcal{O}_{1}  &  \sim O\left(  \frac{\varepsilon}{R}\right)  ^{\delta
-2}\times\left\{
\begin{array}
[c]{c}%
1+O\left(  R/r\right)  ^{2},\;\text{if }\delta\text{ is odd}\\
O\left(  R/r\right)  ,\;\text{if }\delta\text{ is even}%
\end{array}
\right.  ;\\
\mathcal{O}_{2}  &  \sim O\left(  \frac{\varepsilon}{r_{N}}\right)
^{\delta-2}\times O\left(  r_{N}/r\right)  ^{\delta-1}.
\end{align*}

By comparing the strength of the potentials, we can see that $\chi_{s}$ is
greater than $\varphi_{s}$ if
\begin{equation}
r>\left[  \frac{4}{\beta_{\delta}}\frac{Mc^{2}}{Q^{2}/4\pi\epsilon_{0}a_{0}%
}\left(  \frac{\varepsilon}{a_{0}}\right)  \right]  ^{\frac{1}{\delta-3}%
}\varepsilon.
\end{equation}
For realistic values of $\varepsilon$, the above condition is satisfied
outside the nucleus. So, in this scenario, the gravitational potential
produced by the electromagnetic field is greater than the short-distance term
of the gravitational potential of the nuclear mass, in the exterior region.

In three-dimensional space, this does not happen. Indeed, considering the same
distribution (\ref{u}), we find in the exterior region: $\chi_{(3)}%
=-GE/\left(  c^{2}r\right)  +GQ^{2}/\left(  4\pi\epsilon_{0}c^{2}r^{2}\right)
$, where $E$ is proportional to the energy of the electromagnetic field. The
attractive term of $\chi_{(3)}$ depends on the inverse of the distance $r$,
therefore, it can be incorporated in the three-dimensional gravitational
potential $\varphi_{(3)}$ by absorbing the electromagnetic energy as part of
the rest energy of the system. It follows then that the potential,
$\varphi_{(3)}=-GM/r$, with $M=M_{0}+$ $E/c^{2},$ is greater than the
repulsive part of the potential $\chi_{(3)}$ outside the nucleus.

At this point, let us mention that the mass distribution of the nucleus could
be described by a continuous density $\rho_{m}$, instead of a compact
distribution, without changing our conclusion. In this new configuration, the
potential $\varphi_{s}$ would diverge everywhere in the thin brane limit, but
the "dangerous" term would be proportional to $\hat{G}_{D}\rho_{m}%
/\sigma^{\delta-2}$ and still smaller than $\chi_{s}$ for $r>>$ $r_{N}$,
provided that $\rho_{m}$ is a fast decreasing function like an exponential.

In this section, we have calculated the gravitational field produced by the
nucleus. In order to obtain finite results, it was necessary to take into
account the distribution of the nuclear mass and the electromagnetic field
inside the brane. In the next sections, we intend to discuss the complementary
aspect of the picture, namely, the influence of gravity on the dynamics of the
electromagnetic and the Dirac fields in this scenario.

\section{The electrostatic potential in the brane}

In length scales above the brane thickness, the standard model fields can be
treated as traditional four-dimensional fields confined in the brane. Thus, we
may assume that, in this scale, the fields do not couple to the bulk geometry
directly, but their dynamic in the spacetime is influenced by gravity through
the brane geometry. Admitting that the brane is located at $z=0,$ in the given
coordinate system, the induced metric can be directly obtained from
(\ref{Dmetric}). After a transformation to isotropic coordinates, the induced
metric can be written in the following form:%
\begin{equation}
ds^{2}=-w^{2}\left(  dx^{0}\right)  ^{2}+v^{2}\left(  d\vec{x}\cdot d\vec
{x}\right)  , \label{metric}%
\end{equation}
where%
\begin{align}
w^{2}  &  =1+\frac{2}{c^{2}}\varphi_{s}+\frac{2(2+\delta)}{c^{2}(1+\delta
)}\chi_{s},\label{w}\\
v^{2}  &  =1-\frac{2}{c^{2}(1+\delta)}\varphi_{s}+\frac{2(2+\delta)}%
{c^{2}(1+\delta)}\left[  \lambda_{2,s}+\int\frac{\lambda_{2,s}-\lambda_{1,s}%
}{r}dr\right]  . \label{v}%
\end{align}

In this context, let us determine the modification on the electrostatic
interaction between the electron and nucleus caused by the geometry of the
brane. The Maxwell equation in the curved space can be written as%
\begin{equation}
\frac{1}{\sqrt{-g}}\frac{\partial}{\partial x^{\mu}}\left(  \sqrt{-g}F^{\mu
\nu}\right)  =\mu_{0}J^{\nu}, \label{maxwell}%
\end{equation}
where $g$ is the determinant of the metric tensor (\ref{metric}), $J^{\nu}$ is
the four-current and $\mu_{0}$ is the vacuum magnetic permeability. If
$U^{\mu}=dx^{\mu}/d\tau$ is the four-velocity field of the source, then
$J^{\mu}=-\rho_{e}U^{\mu}$, where $\rho_{e}$ is the proper charge density and
$\tau$ is the proper time. For the sake of simplicity, let us admit that the
nucleus is at rest in the given frame. Thus, the normalized four-velocity has
the following components: $U^{\mu}=\left(  cw^{-1},0\right)  ,$ in the given
coordinate system.

In the static regime, the four-potential reduces to $A_{\mu}=\left(
-\phi/c,0\right)  $. Considering that $F_{\mu\nu}=\partial_{\mu}A_{\nu
}-\partial_{\nu}A_{\mu}$, we can derive from (\ref{maxwell}) the field
equation for the electric potential $\phi$:
\begin{equation}
\partial_{i}\left(  \frac{v}{w}\partial^{i}\phi\right)  =-\frac{\rho_{e,0}%
}{\epsilon_{0}},\label{electrostatic}%
\end{equation}
where $\rho_{e,0}=v^{3}\rho_{e}$ is the charge density with the flat measure
$d^{3}x$. This means that in a hypersurface orthogonal to $U^{\mu}$, it
satisfies the condition: $\rho_{e,0}\left(  d^{3}x\right)  =\rho_{e}\left(
v^{3}d^{3}x\right)  $.

As the gravitational field is weak, we expect that the gravitationally
modified electric potential can be written as $\phi=\phi_{0}+\phi_{G}$, where
$\phi_{0}$ is the solution in the flat space sourced by $\rho_{e,0}$ and
$\phi_{G}$ is a small correction of the order of $G_{D}$ due to the spacetime
curvature. In the first order of $G_{D}$, the field equation
(\ref{electrostatic}) reduces to the form%
\begin{equation}
^{0}\nabla^{2}\phi_{G}=-\frac{\rho_{P}}{\epsilon_{0}},
\end{equation}
where $^{0}\nabla^{2}$ is the Laplacian in the flat three-dimensional space
and
\begin{equation}
\rho_{P}=\nabla\cdot\left[  \left(  1-v/w\right)  \epsilon_{0}\vec{E}\right]
\end{equation}
plays the role of a polarization charge density which works here as a source
of the gravitational correction of the electrostatic potential. The solution
is%
\begin{equation}
\phi_{G}\left(  x\right)  =\frac{1}{4\pi\epsilon_{0}}\int\frac{\rho_{P}\left(
x^{\prime}\right)  }{\left\vert \vec{x}-\vec{x}^{\prime}\right\vert }%
d^{3}x^{^{\prime}},
\end{equation}
which is proportional to $\hat{G}_{D}MQ^{2}/c^{4}$ in the leading order.

\section{Dirac equation in the brane}

Now let us discuss how the Hamiltonian of the atom is modified when we take
into account the gravitational interaction of the nucleus with the test
particle. Following the same procedure of the previous section, we shall
assume that the confined fermions do not feel the geometry of the bulk, but
instead they interact with the metric of the $(3+1)$-spacetime given
in\ (\ref{metric}).

Associated with that metric $g_{\mu\nu}$, we find a set of four orthonormal
vector fields, whose components, $e_{\hat{a}}^{\mu}\left(  x\right)  $ (the
vierbein fields), satisfy the equations $g_{\mu\nu}e_{\hat{a}}^{\mu}e_{\hat
{b}}^{\nu}=\eta_{\hat{a}\hat{b}}$. Here the indices $\hat{a}$ and $\hat{b}$
identifies the vector fields and $\mu$ and $\nu$ refer to their components
with respect to the given coordinate system. The indices run from $0$ to $3$.
From the four Dirac matrices ($\gamma^{a}$) defined in the Minkowski space and
with the help of the vierbein fields, we can construct the matrices
$\gamma^{\mu}\left(  x\right)  =e_{\hat{a}}^{\mu}\left(  x\right)  \gamma^{a}$
which satisfy the anti-commutation relation $\left\{  \gamma^{\mu}\left(
x\right)  ,\gamma^{v}\left(  x\right)  \right\}  =2g^{\mu\nu}\left(  x\right)
$ in the curved space.

Assuming a minimal coupling with the gravitational field, the Dirac equation
(in a curved space) that describes the state of a fermion with mass $m$ and
electric charge $q$ subject to gauge field $A_{\mu}$ is given by:%
\begin{equation}
\left[  i\gamma^{\mu}\left(  x\right)  D_{\mu}-mc/\hbar\right]  \psi\left(
x\right)  =0,
\end{equation}
where the operator $D_{\mu}=\nabla_{\mu}-iqA_{\mu}$ and\emph{ }$\nabla_{\mu}$
is the covariant derivative of the spinor, which can be written in terms of
the spinorial connection $\Gamma_{\mu}\left(  x\right)  $ as%
\begin{equation}
\nabla_{\mu}\psi\left(  x\right)  =\left[  \partial_{\mu}+\Gamma_{\mu}\left(
x\right)  \right]  \psi\left(  x\right)  . \label{MP02}%
\end{equation}
On its turn, $\Gamma_{\mu}\left(  x\right)  $ is defined as
\begin{equation}
\Gamma_{\mu}\left(  x\right)  =-\frac{i}{4}\sigma^{ab}e_{\hat{a}}^{\nu}\left(
x\right)  e_{\hat{b}\nu;\mu}\left(  x\right)  , \label{MP03}%
\end{equation}
where $\sigma^{ab}=\frac{i}{2}\left[  \gamma^{a},\gamma^{b}\right]  $ is a
representation of the Lorentz Lie Algebra in the spinor space. As usual, the
symbol $[,]$ is the commutator operator and $e_{b\nu;\mu}\left(  x\right)  $
is the covariant derivative of the vierbein fields which depend on the
Levi-Civita connection $\Gamma_{\nu\mu}^{\lambda}$ (the affine connection
compatible with the metric) according to%
\begin{equation}
e_{\hat{b}\nu;\mu}\left(  x\right)  =\partial_{\mu}e_{\hat{b}\nu}\left(
x\right)  -\Gamma_{\nu\mu}^{\lambda}\left(  x\right)  e_{\hat{b}\lambda
}\left(  x\right)  . \label{MP04}%
\end{equation}

For the diagonal metric (\ref{metric}), a possible choice for the vierbein
fields are:%
\begin{align}
e_{\hat{0}}^{0}\left(  x\right)   &  =w^{-1},\\
e_{\hat{\jmath}}^{i}\left(  x\right)   &  =\delta_{j}^{i}v^{-1},
\end{align}
and all the other components equal to zero.

By a direct calculation, the Dirac equation can be written in the form
$i\hbar\frac{\partial\psi}{\partial t}=H\psi$. The operator $H$ is the atomic
Hamiltonian, which, in a convenient representation, assumes the following form
in the first order of $G_{D}$ \cite{RN Dirac}:%
\begin{equation}
H=\frac{1}{2}\left\{  \vec{\alpha}\cdot\vec{p},\frac{w}{v}\right\}  +w\beta
mc^{2}+q\phi, \label{H}%
\end{equation}
Here $\vec{p}$ is the usual three-dimensional momentum operator in flat
spacetime, $\alpha^{i}=$ $\gamma^{0}\gamma^{i}$ and $\beta=\gamma^{0}$. For
the sake of simplicity, we are neglecting the effects of the potential vector
$\vec{A}$. It is also important to mention that, in the chosen representation,
the Hamiltonian (\ref{H}) is Hermitian in the Hilbert space of
square-integrable functions endowed with the usual inner product calculated
with the flat measure $d^{3}x$.

Following the Foldy-Wouthuysen procedure, the non-relativistic limit of the
Hamiltonian (\ref{H}) can be obtained, by admitting that the mean value of
each term of $H$ is much smaller than the rest energy of the test particle,
$mc^{2}$, in Rydberg states. In the gravitational sector of the Hamiltonian,
i.e., in the part of $H$ constituted by terms proportional to $G_{D}$, we find
that the leading terms are proportional to the test particle mass. Thus:%
\begin{equation}
H_{G}\approx m\varphi_{s}+\frac{2(2+\delta)}{c^{2}(1+\delta)}m\chi_{s}.
\end{equation}
As $\chi_{s}$ is greater than $\varphi_{s}$ in the exterior region, then we
can consider a further approximation:%
\begin{equation}
H_{G}\approx\frac{2(2+\delta)}{c^{2}(1+\delta)}m\chi_{s},\label{HG}%
\end{equation}
in order to find corrections on the energy levels of Rydberg states, coming
from the short-distance behavior of the higher-dimensional gravitational field
produced by the energy of the electrovacuum surrounding the test particle in a
Hydrogen-like ion.

\section{Results and Discussion}

Treating $H_{G}$ as a small term of the total Hamiltonian, we can use the
perturbation method to estimate the gravitational potential energy of a
Hydrogen-like ion in a Rydberg state. For an ion with an atomic number $Z$
found in a state whose principal number is $n$ and $l$ is the angular
momentum, the mean gravitational potential energy is approximately given by%

\begin{equation}
\left\langle n,l\right\vert H_{G}\left\vert n,l\right\rangle =-\hat{\beta
}_{\delta}\gamma_{n,l}\frac{G_{D}Z^{4}e^{2}m}{c^{2}\varepsilon^{\delta-2}%
a_{0}^{4}\epsilon_{0}}, \label{EG}%
\end{equation}
where $e$ is the fundamental electric charge and%
\begin{align}
\hat{\beta}_{\delta}  &  =\frac{\delta\Gamma\left(  \frac{\delta-2}{2}\right)
}{4\pi^{\left(  2+\delta\right)  /2}},\nonumber\\
\gamma_{n,l}  &  =\frac{\left(  3n^{2}-l\left(  l+1\right)  \right)  }%
{n^{5}l(l+1)\left(  2l+3\right)  (2l+1)\left(  2l-1\right)  }.
\end{align}
Actually, it is important to stress, the energy (\ref{EG}) is provided by the
short-distance behavior of the Green function for the gravitational potential.

Without extra-dimensions, the gravitational potential energy of the ion in the
same state is%
\begin{equation}
\left\langle n,l\right\vert H_{G}^{(3)}\left\vert n,l\right\rangle =-\frac
{AZ}{n^{2}}\frac{GmM_{p}}{a_{0}}, \label{EG3}%
\end{equation}
where $A$ is the mass number. Here for the sake of simplicity, we admit that
the mass of the nucleus is approximately equal to $AM_{p}$.

Comparing (\ref{EG}) and (\ref{EG3}), we can find conditions under which the
amount of gravitational potential energy coming from the short-distance
behavior of the higher-dimensional potential surpasses the three-dimensional
value. In this domain, extra dimensions may amplify significantly the
gravitational effects on the ion. The condition $\left\langle H_{G}%
\right\rangle >\left\langle H_{G}^{(3)}\right\rangle $ can be expressed in
terms of the size of the extra dimensions $R$ in the following form:
\begin{equation}
\left(  2\pi R\right)  ^{\delta}>\frac{1}{\hat{\beta}_{\delta}\gamma_{n,l}%
}\frac{A}{Z^{3}n^{2}}\frac{\epsilon_{0}M_{p}c^{2}a_{0}^{3}\varepsilon
^{\delta-2}}{e^{2}}%
\end{equation}
Notice that, in order to the gravitational potential energy be amplified by
the extra-dimension, $R$ should be some orders of magnitude greater than the
brane thickness, but interestingly the compactification scale can be smaller
than the Bohr radius. As an example, consider the Neon ion ($^{20}$Ne$^{+9}$)
in a six-dimensional space. In a state with $n=15$ and $l=14$, the extra
dimensions would provide an amplification if $R>10^{-14}$ $%
\operatorname{m}%
$, when $\varepsilon=10^{-20}$ $%
\operatorname{m}%
$, for instance.

Now let us consider the transition between the states $(n,l=n-1)$ and
$(n-1,l=n-2)$. The principal part of the energy gap between these levels is%
\[
\Delta E_{P}=\frac{Z^{2}e^{2}}{8\pi\epsilon_{0}a_{0}}\left(  \frac{1}{n^{2}%
}-\frac{1}{\left(  n-1\right)  ^{2}}\right)  .
\]
From (\ref{EG}), we can calculate the difference of the gravitational
potential energy between the mentioned levels ($\Delta E_{G}$). Relative to
the principal part of the energy gap, we find
\begin{equation}
\frac{\Delta E_{G}}{\Delta E_{P}}=-\hat{\beta}_{\delta}\frac{1}{8\pi}%
\frac{\left(  \gamma_{n,n-1}-\gamma_{n-1,n-2}\right)  }{n^{-2}-\left(
n-1\right)  ^{-2}}\frac{Z^{2}G_{D}m}{c^{2}\varepsilon^{\delta-2}a_{0}^{3}}.
\label{relativeEG}%
\end{equation}

If we consider muonic hydrogen-like ions, then the Bohr radius of the
ground-state of the Hydrogen, $a_{0}$, is replaced by the Bohr radius of the
muonic Hydrogen, $a_{0,\mu}$, and the test particle mass $m$ correspond to the
muon mass $m_{\mu}$. Due to this, (\ref{relativeEG}) is multiplied by a factor
$\left(  a_{0}/a_{0,\mu}\right)  ^{3}\left(  m_{\mu}/m\right)  \sim$ $10^{9}$.

According to Ref. \cite{Jentschura}, if the frequency transition lies in the
optical band, the relative experimental precision could reach the order of
$10^{-19}$ \cite{opticalmetrology}. Given this fantastic precision, we are led
to think about the possibility of using Rydberg states in the search for
hidden dimensions. In a previous paper \cite{Dahia2}, it was suggested that
the proton radius puzzle can be explained in the brane scenario, once the
muon-proton gravitational interaction, modified by extra-dimensions, can
account for the unexpected energy excess measured in the muonic Hydrogen Lamb
shift. To solve the proton radius puzzle, the fundamental Planck mass $M_{D}$,
or equivalently, the fundamental gravitational constant $G_{D}$, should have a
certain value given in terms of the parameter $\sigma$, which here can be seen
as the confinement parameter of the nucleus in the brane. Taking this value of
$G_{D}$ as reference in equation (\ref{relativeEG}) and admitting that the
matter and the electromagnetic fields have the same confinement parameter
$(\varepsilon\simeq\sigma)$, we show in Figure 1 what are the optical
transitions between Rydberg states in which the effect of the
higher-dimensional gravity exceeds the experimental precision promised by the
optical metrology \cite{opticalmetrology}.%
\begin{center}
\includegraphics[
height=1.9086in,
width=3.2785in
]%
{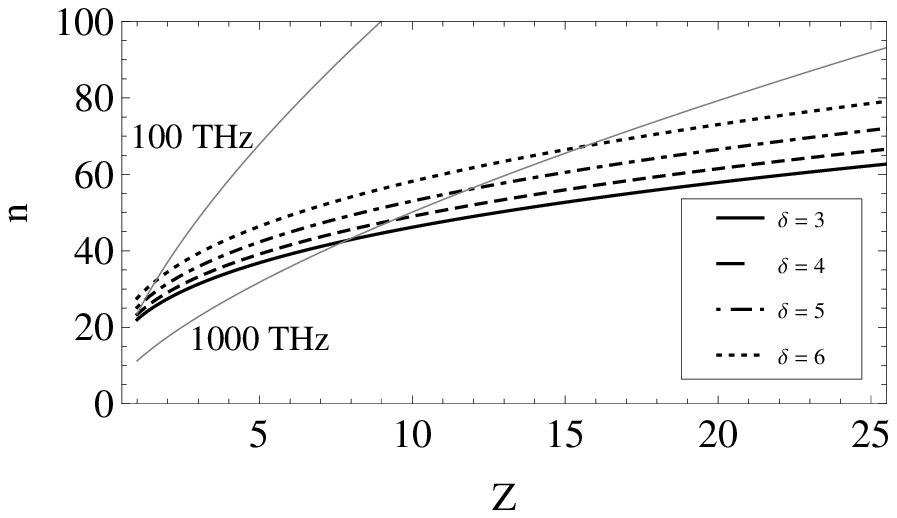}%
\\
In transitions between Rydberg states of muonic Hydrogen-like ions below the
dashed lines, the influence of hidden dimensions, through the classical
higher-dimensional gravity, can be greater than promised experimental
uncertainty in measurements of transition frequencies around the optical band
(100 THz -1000 THz). Here we are assuming that $G_{D}$ has the appropriate
value to solve the proton radius puzzle \cite{Dahia2}.
\label{Figure1}%
\end{center}

From this result, we see that for some muonic-hydrogen like ions the
spectroscopy of Rydberg states in the optical frequency could, in principle,
be employed in order to search for traces of hidden dimensions and, in
particular, to test the hypothesis according to which the proton radius puzzle
could be explained by the existence of extra dimensions \cite{Dahia2}.
However, it is important to remark that these predictions are based on the
classical behavior of the gravitational field. On the other hand, as pointed
out in Ref. \cite{colliders}, quantum-gravity effects would become relevant at
a length scale around the fundamental Planck length ($\ell_{D}$) defined in
the higher dimensions or even in greater distances. If this is the case, then
quantum-gravity effects could modify the classical result unpredictably.
Nevertheless, as a definite quantum-gravity theory is not yet known, only
experiments can answer this question.

\section{Final remarks}

The prospect of measuring optical transitions between Rydberg states with a
relative uncertainty near to the impressive order of $10^{-19}$ motivated us
to investigate the influence of hidden dimensions on the energy levels of the
Rydberg states of hydrogen-like ions.

In the braneworld scenario with large extra dimensions, gravity is the first
interaction to be affected by the supplementary space. These models predict
that, in comparison to the three-dimensional Newtonian interaction, the
gravitational field can be hugely amplified in short distance. In this paper,
we have investigated the effects of extra dimensions in the atomic domain by
studying the influence of the higher-dimensional gravitational field produced
by the nucleus on the energy levels of the ion.

Due to the electric charge of the source, the spacetime around the nucleus is
a brane-version of the Reissner-Nordstrom geometry. In this kind of spacetime,
the metric, in the weak field regime, depends on the gravitational potential
produced by the rest energy concentrated on the nucleus, $\varphi$, and also
on the gravitational potential generated by the energy of the electric field
created by the nucleus charge, $\chi$.

In the zero-thickness approximation, the short-distance potential $\chi_{s}$
diverges everywhere in the brane. Due to this, the potential gravitational
energy is not computable in the thin brane limit even when the ion is in
Rydberg states with high angular momentum.

Therefore, in order to estimate the influence of the potential $\chi_{s}$ on
the energy of the ion, we had to address this problem in the thick brane
scenario. Considering a uniform distribution of the electromagnetic energy
inside a brane strip, we calculate the gravitational potential energy in terms
of the confiment parameter of the electric field. We find that the extra
dimensions are capable of amplifying the gravitational potential energy of the
ion in a Rydberg state even when the compactification radius is smaller than
the Bohr radius. This a consequence of the behavior of the potential $\chi
_{s}$.

It is interesting that the short-distance behavior of the Green function
associated to the gravitational potential plays an important role in the
present situation. At the first sight, we could be led to think that the
short-distance behavior is irrelevant when the ion is the Rydberg state, since
the probability to find the test particle close to the nucleus is very small
in this state. Indeed this is the reason why the potential $\varphi_{s}$ is
weak here. However, the source of the potential $\chi$ is spread in the space.
Therefore, around any external point $\vec{x}$, there is an interval smaller
than $R$, in the electrovacuum, where the short-distance Green function is
dominant. It is through this term that the gravitational influence of the
electrovacuum surrounding the test particle can increase significantly the
gravitational potential energy of the ion, to the point of making possible the
use of the spectroscopy of Rydberg states in the search for hidden dimensions.

\begin{acknowledgement}
A. S. Lemos and E. Maciel thanks CAPES for financial support.
\end{acknowledgement}


\begin{thebibliography}{99}                                                                                               %


\bibitem {ADD1}%
N.~Arkani-Hamed, S.~Dimopoulos and G.~R.~Dvali,
Phys.\ Lett.\ B {\bf429}, 263 (1998).%


\bibitem {ADD2}%
I.~Antoniadis, N.~Arkani-Hamed, S.~Dimopoulos and G.~R.~Dvali,
Phys.\ Lett.\ B {\bf436}, 257 (1998).%


\bibitem {RS1}%
L.~Randall and R.~Sundrum,
Phys.\ Rev.\ Lett.\  {\bf83}, 3370 (1999).%


\bibitem {RS2}%
L.~Randall and R.~Sundrum,
Phys.\ Rev.\ Lett.\  {\bf83}, 4690 (1999).%


\bibitem {Hoyle2007}%
D.~J.~Kapner, T.~S.~Cook, E.~G.~Adelberger, J.~H.~Gundlach, B.~R.~Heckel, C.~D.~Hoyle and H.~E.~Swanson,
Phys.\ Rev.\ Lett.\  {\bf98}, 021101 (2007).%


\bibitem {review}Jiro Murata and Saki Tanaka, Class. Quantum Grav. \textbf{32}
033001 (2015).

\bibitem {SN}S. Cullen and M. Perelstein, Phys. Rev. Lett., \textbf{83,} 268 (1999)

\bibitem {neutronstar}S. Hannestad and G. G. Raffelt, Phys. Rev. D
\textbf{67,} 125008 (2003); Erratum-ibid.D, \textbf{69,} 029901 (2004).

\bibitem {colliders}Gian E Giudice, Riccardo Rattazzi and James D. Wells,
Nuclear Physics B, \textbf{544} (1999) 3-38.

\bibitem {lhc}Aad G et al. (Atlas Collaboration), Phys. Rev. Lett.
\textbf{110,} 011802 (2013). CMS Collab., Phys. Lett. B \textbf{755}, 102
(2016) .

\bibitem {monojet}CMS Collab., Eur. Phys. J. C \textbf{75} 235 (2015).

\bibitem {landsberg}Greg Landsberg, Mod. Phys. Lett. A\textbf{50}, 1540017 (2015).

\bibitem {pdgAlex}K.A. Olive et al. (Particle Data Group), Chin. Phys. C,
\textbf{38}, 090001 (2014), \textit{(Extra dimensions, Updated September 2015
by John Parsons and Alex Pomarol).}

\bibitem {specforxdim1}Feng Luo, Hongya Liu, Chin. Phys. Lett. \textbf{23,
}2903, (2006). Feng Luo, Hongya Liu, Int. J. of Theoretical Phys.,
\textbf{46}, 606 (2007).

\bibitem {atomicspec}Li Z-G, Ni W-T and A. P. Pat\'{o}n, Chinese Phys. B,
\textbf{17,} 70 (2008).

\bibitem {Li}%
Z.~Li and X.~Chen,
arXiv:1303.5146 [hep-ph].%


\bibitem {Wang}%
L.~B.~Wang and W.~T.~Ni,
Mod.\ Phys.\ Lett.\ A {\bf28}, 1350094 (2013).%


\bibitem {specforxdim2}Zhou Wan-Ping, Zhou Peng, Qiao Hao-Xue, Open Phys.,
\textbf{13}, 96 (2015).

\bibitem {molecule}E J Salumbides et al, New J. Phys. \textbf{17} 033015 (2015).

\bibitem {dahia}F. Dahia and A.S. Lemos, Phys. Rev. D 94 no.8, 084033 (2016).

\bibitem {Jentschura}Ulrich D. Jentschura, Peter J. Mohr, Joseph N. Tan, and
Benedikt J. Wundt, Phys. Rev. Lett. 100, 160404 (2008).

\bibitem {codata}P.J. Mohr, B.N. Taylor, D.B. Newell, Rev. Modern Phys.
\textbf{84}, 1527 (2012).

\bibitem {opticalcomb}T.W. H\"{a}nsch, Rev. Mod. Phys. 78, 1297 (2006)

\bibitem {opticalmetrology}L.-S. Ma et al., Science 303, 1843 (2004).

\bibitem {nature}R. Pohl et al., Nature \textbf{466,} 213 (2010).

\bibitem {science}A. Antognini et al, Science \textbf{339}, 417 (2013).

\bibitem {carl}Carl E. Carlson, Progress in Particle and Nuclear Physics
\textbf{82}, 59, (2015).

\bibitem {krauth}J. J. Krauth \textit{et al, } arXiv:1706.00696.

\bibitem {effbrane}Francisco del Aguila, Manuel Perez-Victoria, Jose Santiago,
JHEP 0610:056 (2006).

\bibitem {radion}E. G. Adelberger, B.R. Heckel, S. Hoedl, C. D. Hoyle, D. J.
Kapner and A. Upadhye, Phys. Rev. Lett \textbf{98}, 131104 (2007).

\bibitem {kehagias}%
A.~Kehagias and K.~Sfetsos,
Phys.\ Lett.\ B {\bf472}, 39 (2000).%


\bibitem {rubakov}V. Rubakov and M. Shaposhnikov, Phys. Lett. B \textbf{125},
136 (1983).

\bibitem {antoniadis}I. Antoniadis, K. Benakli, A. Laugier, T. Maillard,
Nucl.Phys. B \textbf{662,} 40 (2003).

\bibitem {RN Dirac}J. H. Noble and U. D. Jentschura, Phys. Rev. A 93, 032108 (2016).

\bibitem {Dahia2}F. Dahia and A. S. Lemos, Eur. Phys. J. C76, 8, 435 (2016).
\end{thebibliography}
\end{document}